\documentstyle[12pt]{article}
\newcommand{\LI}{\hbox to\hsize}
\newcommand{\LLI}[1]{\LI{#1\hss}}

\newcommand{\PM}[1]%
{\mbox{$m_{\rm #1}$}} %generic particle mass
\newcommand{\BEQ}{\begin{equation}}
\newcommand{\EEQ}{\end{equation}}
\newcommand{\bra}[1]{\mbox{$\langle #1 \left| \right. $}}
\newcommand{\ket}[1]{\mbox{$ \left| \right.  #1 \rangle$}}

\newcommand{\ack}{\vskip6mm \LLI{\Large{\bf Acknowledgement}}
\vspace*{2mm}
\par
\noindent
\nopagebreak
This research has been supported in part by the U.S. Department of Energy
under Grant \#~DE--FG--02--85ER40211.} % Acknowledges grant support
\newcommand{\incircle}[1]{\mbox{{\hbox{$\bigcirc$}\kern-0.7em
\lower0.05ex\hbox{\mbox{{\scriptsize\rm #1}}}}}}
\newcommand{\eq}[1]{eq.~(\ref{#1})}
\newcommand{\ETC}{\mbox{\em etc.\/ }}
\newcommand{\VIZ}{\mbox{\em viz.\/ }}

\newcommand{\IE}{\mbox{\em i.e. \/}}

\newcommand{\EG}{\mbox{\em e.g.\/ }}

\newcommand{\mathfig}[4]{%
\begin{figure}[#4] % position the figure by #4
\vspace*{6in}
\centerline{\hbox to 14in % center the file and set its width
 {\hskip5in {\special{eps:#2 x=14in}\hfil}}} % cut off a piece of the
\vspace*{-1.5in} % move up the the caption
\caption{#3}
\label{#1}
\end{figure}\vskip0.5in} % leave space between the caption and the next
\begin{document}
%\RLI{JHU--TIPAC 940015}
%\RLI{August 1994}
%\vskip5mm
\begin{center}
{\Large\bf Resonant Neutrino Interaction and Mixing}\\[4mm]
G. Domokos and S. Kovesi--Domokos\\
INFN, Sezione di Firenze\\
Florence, Italy\\
and\\
The Henry A. Rowland Department of Physics and Astronomy\\
The Johns Hopkins University\\
Baltimore, MD 21218\footnote{Permanent address. E--mail:~SKD@JHUP.PHA.JHU.EDU}

\end{center}
{\small We investigate the resonant interaction of neutrinos in matter
in the presence of mixing. At energies near the W resonance,
oscillations are absent;
the survival probability of electron antineutrinos is suppressed due
to the presence of open inelastic channels.}\vskip3mm

The existence of a resonant interaction of electron antineutrinos
in matter and the possibility of flavor oscillations have been known for
a long time; for a summary see~\cite{kimpevsner}. The laboratory energy
at which the resonance occurs is about 6.4~PeV.
Recently,
interest has turned towards an investigation of neutrino oscillations
in the PeV energy range, at energies near the  W resonance in the
$\bar{\nu}_{e}e$ interaction cross section~\cite{learned}.
Learned and Pakvasa in ref.~\cite{learned} point out that neutrino
oscillations in a hitherto unexplored range of neutrino masses and mixing
angles may lead to the arrival and detection of $\tau$ neutrinos in present
and planned neutrino telescopes (DUMAND, NESTOR, KM3 \ETC).

It is generally assumed~\cite{hawaii} that neutrinos emerging from a
celestial source
(typically, an
AGN or a binary system) are produced as a result of the decay of hadrons.
If so, in either hadron--hadron interactions or in
photoproduction most of the hadrons produced are pions, with charged
pions of either sign being produced in approximately equal numbers and
energy distributions. Consequently, the number of neutrinos and
antineutrinos of any flavor is approximately equal.
To a reasonable approximation, the ratio of neutrinos
(antineutrinos, {\em resp\/.}) of different flavors produced is:
\[
\nu_{e}:\nu_{\mu}:\nu_{\tau}\approx 1:2:0
\]

It has to be noted, however that at least some of the time,
a considerable amount of matter is present between the primary source
of neutrinos and the detectors in
the experiments mentioned.
\begin{itemize}
\item In all probability, neutrinos
produced in an AGN or a binary system, have to pass through a region
of high electron (and
perhaps baryon) density.

\item In addition, the neutrinos usually pass through a certain amount of
matter between the source and the detector, \EG
the Earth (in the case of the observation of upward going neutrinos) and/or
sea water (in case of zenith angles close to the horizon).
\end{itemize}

In this note we begin to investigate the question
 of neutrino oscillations if neutrinos
in the PeV energy range (around the energy of where the W resonance occurs)
penetrate
matter between the source and the detector. Our main purpose here is
to explore the qualitative aspects of the phenomena. For this reason,
we use the formalism developed by Wolfenstein~\cite{wolfenstein}
and a hypothetical situation of having two flavors only.  Furthermore,
the medium
is assumed to be a homogeneous one.

A full
treatment taking into account all three flavors and a more complete
treatment of  the presence of inelastic channels
and inhomogeneities in the medium will be given elsewhere.

The Wolfenstein equation can be viewed as a Schr\"{o}dinger equation
 in a null plane formalism.
(It has been known for a long time that the kinematics in the
two dimensional transverse space is Galilean; for a description
of the formalism with an emphasis on its
  group theoretical aspects,
see~\cite{nullplane, boulder}, particularly the second of these
references.)

The behavior of the evolution operator of neutrinos, $E$, is
governed by the equation:
\BEQ
i\partial_{t}E =  \frac{\left({\stackrel{\rightarrow}{P}}\hskip0.1em ^{2}
+  M^{2} +V\right)}{2p}E,
\label{eq:wolfenstein}
\EEQ
where $p$ and $\stackrel{\rightarrow}{P}$ stand for the longitudinal
and transverse
components of the momentum, respectively; $M^{2}$ and $V$ are the
mass squared and interaction operators. The variable $t$ may be thought
of as the time conjugate to the Hamiltonian,
\[ H = \frac{\stackrel{\rightarrow}{P}\hskip0.1em ^{2} + M^{2} + V}{2p}. \]

The evolution operator as
well as the mass squared, interaction and transverse momentum operators
are to be thought of as $f\times f$ dimensional matrices, where $f$ is
the number of flavors. The operator $\stackrel{\rightarrow}{P}$ is
proportional to
the unit matrix.
The important feature of the null plane formalism
is that the longitudinal and transverse degrees of freedom are almost
exactly separated. Thus, $p$ may be diagonalized separately and quantities
like $V$, governing the dynamics in the transverse plane may
depend on its eigenvalue. (For the same reason, $p$ is,
to a good approximation, equal
to the kinetic energy of the neutrinos at the source.)

In the conventional treatment of neutrino oscillations (for a review,
see~\cite{kuopantaleone}) one omits the square of the transverse
momentum and represents the matter present in terms of a potential
independent of the transverse degrees of freedom. (This is equivalent to
the classic Lorentz formula, expressing the polarizability of a medium
in terms of the forward scattering amplitude of light on the particles
in the medium.)

Since we are to treat a resonant interaction, such an approximation is
not an adequate one. We can still neglect  the transverse degrees
of freedom; however, a more accurate expression has to be found for the
``potential''. The appropriate  procedure has been known
for a long time, see~\cite{brueckner} and the original
references quoted there.

Briefly, the recipe is that
wherever a potential appears in the  the lowest order treatment
of the scattering, it has to be replaced by an {\em effective
potential\/}. The latter is given by the level shift operator
(the ``$K$ matrix'')
of the scattering of the propagating particle in the medium. The
relationship between the transition operator, ${\bf T}$ and the ${\bf K}$
matrix is:
\BEQ
{\bf T} = - {\bf K} \left( 1+i{\bf K}\right)^{-1}
\label{eq:kmatrix}
\EEQ

We refer the reader to Brueckner's lecture in ref.~\cite{brueckner}
for a derivation of the ``recipe''just quoted. However, we can
make it  plausible by recalling that the $K$ matrix obeys
the Lippmann--Schwinger equation. In an operator form the latter reads:
\BEQ
{\bf K} = {\bf V} + {\bf V} \frac{{\cal{P}}}{\epsilon - {\bf H}_{0}}{\bf K},
\label{eq:lippmannschwinger}
\EEQ
where $\epsilon$ is the energy (in the present context, $\epsilon = p$)
and ${\bf H}_{0}$ is the free Hamiltonian,
see \EG ref.~\cite{goldbergerwatson}.
The symbol of the principal
value, $\cal{P}$, simply means that  those eigenstates of the free
Hamiltonian for which the inverse of $\epsilon - H_{0}$ is singular,
are to be deleted from a complete set. One now realizes that the first
term in the iterative solution of \eq{eq:lippmannschwinger} is indeed the
potential; moreover, in the absence of open inelastic channels, the
operator ${\bf K}$ is a Hermitean one, thus it qualifies as a ``potential''
in a Schr\"{o}dinger equation.

Armed with this knowledge, we can now construct a phenomenological effective
potential describing the resonant interaction of electron antineutrinos
in matter. We describe the scattering amplitude (the matrix element of
 ${\bf T}$) by means of a Breit--Wigner
formula; the forward amplitude is given by:
\BEQ
T= \frac{s {\rm M}^{2}G_{F}\sqrt{2}/6 \pi}{{\rm M}^{2} - s
- i{\rm M}\Gamma_{t}}.
\label{eq:breitwigner}
\EEQ
In the last equation, $M$ stands for the mass of the W, $\Gamma_{t}$ is
its total width. $G_{F}$ stands for the conventional Fermi coupling
constant and $s$ is the square of the center of mass energy. In terms
of the mass of the electron and $p$ it is given by
$s=2m_{\rm e}p$. We note that in terms of these quantities, the elastic width
of the W is given by the expression, $\Gamma_{e} = M^{3}G_{F}/6 \pi \sqrt{2}$.
In what follows, we take the numerator of \eq{eq:breitwigner} at $s= M^{2}$;
this is a good approximation for a narrow resonance.
Due to the fact that we neglect all but the resonant partial wave
in \eq{eq:breitwigner}, the expression of the matrix element of the ${\bf K}$
matrix
is found immediately, \VIZ
\BEQ
K = -\frac{M \Gamma_{e}}{{\rm M}^{2} - s -iM\Gamma_{i}},
\label{eq:reskmatr} \EEQ
where $\Gamma_{i} = \Gamma_{t} - \Gamma_{e}$ is the inelastic width
of the W.
If the inelastic width, $\Gamma_{i} =
\Gamma_{t} - \Gamma_{e}$ were zero, the expression in \eq{eq:reskmatr}
would be real as required. Thus the effective potential in the medium entering
\eq{eq:wolfenstein} is given by the expression,
\BEQ
V_{eff} = \frac{n_{\rm e}}{m_{\rm e}} K,
\label{eq:effpotential}
\EEQ
where $n_{\rm e}$ is the electron density in the medium.

In the present model, the flavor space is two dimensional; we denote
the flavors by $\rm e$ and $\mu$, respectively. Due to the fact that
(in real life), both the electron and muon are much lighter than the W,
the matrix elements of the effective potential, $\cal{F}$ are:
\BEQ
\bra{\mu}{\cal{F}}\ket{{\rm e}} =\bra{{\rm e}}{\cal{F}}\ket{{\rm e}} = F
\label{eq:F}
\EEQ
and all other matrix elements vanish.

Next, we give the matrix elements of the mass squared matrix, ${\cal{M}}^{2}$:
\begin{eqnarray}
\bra{e}{\cal{M}}^{2}\ket{{\rm e}} & = & - \bra{\mu}{\cal{M}}^{2}\ket{\mu}
\nonumber \\
\mbox{}  &  = & \Delta ^{2}\cos 2 \Theta \nonumber \\
\bra{\mu}{\cal{M}}^{2}\ket{{\rm e}} & = &\bra{{\rm e}}{\cal{M}}^{2}\ket{{\rm
e}}
\nonumber \\
\mbox{} & = & \Delta^{2} \sin 2 \Theta
\label{eq:massmatrix}
\end{eqnarray}
As usual, we subtracted the average mass squared; its presence merely
multiplies the state vectors by an overall phase factor, hence, it has no
influence on neutrino oscillations.

We now have to solve \eq{eq:wolfenstein} with the effective Hamiltonian
put together from eqs.~(\ref{eq:massmatrix}) and (\ref{eq:F}). Despite the fact
that the Hamiltonian is not a
Hermitean matrix in flavor space, the
equation can be solved immediately. This is due to the fact that
any non singular $2\times 2$ matrix can be diagonalized by means of a
similarity transformation. (If the matrix in question is not a Hermitean
one, the diagonalizing matrices are elements of the group SL(2,C) instead of
SU(2).) We merely quote the result of the calculation.
\vskip3mm
\begin{minipage}{\hsize}
\begin{eqnarray}
\bra{{\rm e}}E\ket{{\rm e}} & = & {\rm exp}\left(-it\frac{F}{4p}\right)\left[
\cos \phi - i \sin \phi \frac{\Delta^{2} \cos 2\Theta + F/2}{\mu^{2}}
\right]
\nonumber \\
\bra{\mu}E\ket{mu} & = &{\rm exp}\left( -it\frac{F}{4p}\right)\left[
\cos \phi +i \sin \phi \frac{\Delta^{2} \cos 2\Theta +F/2}{\mu^{2}}\right]
\nonumber \\
\bra{\mu}E\ket{{\rm e}} & = &-i {\rm exp}\left( -it \frac{F}{4p}\right)
\sin \phi \frac{\Delta ^{2} \sin 2\Theta + F}{\mu ^{2}}\nonumber \\
\bra{e}E\ket{\mu} & = & -i {\rm exp}\left( -it \frac{F}{4p}\right)
\sin\phi \frac{\Delta^{2}\sin 2\Theta}{\mu^{2}}
\label{eq:evolution}
\end{eqnarray}
\end{minipage}
\vskip3mm
The quantities $\mu^{2}$ and $\phi$ entering these equations are
defined as follows.
\begin{eqnarray}
\mu^{2} & = & \left( \Delta^{4} + F \Delta^{2}(\cos 2\Theta
+ \sin 2\Theta) + F^{2}/4\right)^{1/2}\nonumber \\
\phi & = &\frac{t\mu^{2}}{2p}\nonumber
\end{eqnarray}

The physical interpretation of these equations is not quite easy,
due to the fact that the quantity F is complex. However, near
the resonance itself (where the use of the Breit--Wigner formula is
justified in the first place), considerable simplifications occur.
In order to understand this,
we notice that the quantity $n_{\rm e}/m_{\rm e}$ ocurring in the
definition of the effective potential, defines a mass scale characterizing
the medium. In the following Table we exhibit approximate values of the
mean electron density, together with the characteristic mass scale,
\mbox{${\rm m}_{c}^{2} = n_{\rm e}/m_{\rm e}$} for some environments
of interest.
\vskip4mm
\begin{minipage}{\hsize}
\begin{center}
{\bf Electron densities and characteristic masses \\ for some
environments}\\[3mm]
\begin{tabular}{|c|c|c|}\hline
Environment & $n_{\rm e} [{\rm cm}^{-3}]$ &${\rm m}_{c}^{2} [{\rm eV}^{2}]$\\
\hline
stellar interior (sun) & $10^{27}$ & $2\times 10^{7}]$\\
Earth & $1.6\times 10^{24}$ & $ 3\times 10^{4} $\\
water & $3\times 10^{23}$ & $5\times 10^{3}$\\ \hline
\end{tabular}
\end{center}
\end{minipage}
\vskip4mm

We see that the characteristic mass scales are, in general, considerably
larger than the expected values of $\Delta^{2}$ in a typical mixing
scheme occurring in the literature, see, for instance,~\cite{kimpevsner}
for a review. As a consequence, the expression of $\mu^{2}$ is
considerably simplified near $s=M^{2}$. In fact, we have:
\BEQ
\mu^{2} \approx \frac{F}{2} + \Delta^{2}\left( \sin 2\Theta + \cos
2\Theta\right),
\label{simplemu}
\EEQ
so that to leading order, $\mu^{2}\approx F/2$.
Using the  Euler decomposition of the trigonomet\-ric
func\-tions, one re\-al\-izes that in the terms pro\-por\-tional to
\[{\rm exp}\left(-itF/4p\right){\rm exp} i\phi , \]
$F$ cancels out to a good approximation. By contrast,
in terms proportional to
\[ {\rm exp}\left(-itF/4p\right){\rm exp}- i\phi , \]
$F$ is present in the exponent. The quantity $F$ has an imaginary part
due to the fact that there are {\em inelastic channels open}: W decays mostly
hadronically. Therefore terms proportional to ${\rm exp}( -i Ft/2p)$
are exponentially damped and all surviving terms in the expression of
the transition matrix have the same phase. (By putting in numbers
one concludes that the extinction distance is a few metres at energies
near the resonance in any environment of interest;
thus, at any reasonable value of $\Delta^{2}$
the damped terms are negligible.) Consequently,
we omit terms proportional to $\exp( - {\rm Im}F t)$ in what follows.

We now use the approximation introduced in \eq{simplemu} to give the
expressions of the transition probabilities. This is best done by
introducing the dimensionless distance in energy from the resonance,
$x$
and an appropriately scaled ratio of $\Delta^{2}/m_{c}^{2}$:
\begin{minipage}{\hsize}
\begin{eqnarray}
x & = &\frac{s - M^{2}}{M \Gamma_{i}} \nonumber \\
\delta & = & \frac{\Delta^{2}\Gamma_{i}}{m_{c}^{2}\Gamma_{e}}
\end{eqnarray}
\end{minipage}
\vskip3mm
The transition probabilities are then given by the expressions:
\vskip2mm
\begin{minipage}{\hsize}
\begin{eqnarray}
\left| \bra{e} E \ket{e}\right|^{2} & = & \frac{\left( \delta
\sin 2 \Theta\right)^{2}}{1+ \left( x + 2\delta \left(\sin 2\Theta
+ \cos 2\Theta\right)
\left( 1 + x^{2}\right)\right)^{2}} \nonumber \\
\left| \bra{\mu} E \ket{e}\right|^{2} & = & \frac{ 1 + \left(
x + \delta \sin 2\Theta \left(1+ x^{2}\right)\right)^{2}}{1+ \left( x
+ 2\delta \left(\sin 2\Theta + \cos 2\Theta\right)
\left( 1 + x^{2}\right)\right)^{2}}\nonumber \\
\left| \bra{\mu} E \ket{\mu}\right|^{2} & = & \frac{ 1 + \left(
x + \delta \sin 2\Theta \left(1+ x^{2}\right)\right)^{2}}{1+ \left( x
+ 2\delta \left(\sin 2\Theta + \cos 2\Theta\right)
\left( 1 + x^{2}\right)\right)^{2}}\nonumber \\
\left| \bra{e} E \ket{\mu}\right|^{2} & = &\frac{ \left(\delta
\sin 2\Theta \right)^{2}\left( 1+ x^{2}\right)^{2}}{1+ \left( x
+ 2\delta \left(\sin 2\Theta + \cos 2\Theta\right)
\left( 1 + x^{2}\right)\right)^{2}}
\label{probabilities}
\end{eqnarray}
\end{minipage}
\vskip3mm
By looking at the last set of equations, one realizes that the survival
probability of electron antineutrinos is very low; this is due to
the fact that the W resonance is
a highly inelastic one: once the electron antineutrino interacts,
the W decays into the original channel only rarely. By contrast,
the muon antineutrinos are hardly depleted and they hardly ever produce
electron antineutrinos: these processes proceed through  mixing only.
Remarkably, the $t$--dependence of all transition probabilities
disappears beyond the extinction distance: an inelastic resonance
produces uniform transition probabilities.

Let us summarize.
\begin{itemize}
\item The presence of an {\em inelastic} resonance
stongly suppresses oscillation effects.
\item The original
1:2:0 ratio of the antineutrinos of the three flavors is strongly distorted.
Electron
antineutrinos are depleted, while muon
antineutrinos are hardly affected. (In fact, their number is slightly
increased due to the decay ${\rm W} \rightarrow \mu \bar{\nu}_{\mu}$.)
\item One can reasonably conjecture that in a calculation
taking three generations into account, $\bar{\nu}_{\tau}$
production will be comparable to the $\bar{\nu}_{\rm e}\rightarrow
\bar{\nu}_{\mu}$ rate (with the appropriate mixing angles and mass
differences), \IE a considerable  number of $\tau$ leptons and the
corresponding antineutrinos will be produced. In particular,
the appearence of $\bar{\nu}_{\tau}$-s in a neutrino
telescope does not necessarily imply oscillation lengths comparable
to the distance to some AGN.
\item The ratios
of neutrinos (as opposed to antineutrinos) should be
unaffected unless there exist neutrino--antineutrino
oscillations.
\end{itemize}
\ack \vskip1mm
We also thank Roberto Casalbuoni, John Learned and Leon Madansky for
enlightening conversations on the
subject.

Most of  this work has been done during the authors' visit at the
Istituto di Fisica, Universit\'{a} di Firenze, during the Summer of 1994.
We thank Roberto Casalbuoni for his hospitality and the INFN for partial
financial support.

\end{document}